\documentstyle[multicol,aps,pra,epsf,array]{revtex}
\begin{document}
\newcommand{\be}{\begin{equation}}
\newcommand{\ee}{\end{equation}}
\newcommand{\bq}{\begin{eqnarray}}
\newcommand{\eq}{\end{eqnarray}}
\newcommand{\Sc}{Schr\"odinger\,\,}
\newcommand{\Sp}{\,\,\,\,}
\newcommand{\no}{\nonumber\\}
\newcommand{\tr}{\text{tr}}
\newcommand{\p}{\partial}
\newcommand{\la}{\lambda}
\newcommand{\La}{\Lambda}
\newcommand{\G}{{\cal G}}
\newcommand{\D}{{\cal D}}
\newcommand{\E}{{\cal E}}
\newcommand{\W}{{\bf W}}
\newcommand{\de}{\delta}
\newcommand{\al}{\alpha}
\newcommand{\bi}{\beta}
\newcommand{\ep}{\varepsilon}
\newcommand{\ga}{\gamma}
\newcommand{\epp}{\epsilon}
\newcommand{\vep}{\varepsilon}
\newcommand{\th}{\theta}
\newcommand{\om}{\omega}
\newcommand{\si}{\sigma}
\newcommand{\J}{{\cal J}}
\newcommand{\pr}{\prime}
\newcommand{\ka}{\kappa}
\newcommand{\TH}{\mbox{\boldmath${\theta}$}}
\newcommand{\DE}{\mbox{\boldmath${\delta}$}}
\newcommand{\lan}{\langle}
\newcommand{\ran}{\rangle}
\newcommand{\Hol}{\text{Hol}}
\newcommand{\cp}{{\bf CP}}
\newcommand{\spp}{\,\,\,\,\,\,\,\,\,\,\,\,\,\,}

\title{Geometric phases of mesoscopic spin in
Bose-Einstein condensates}
\author{I. Fuentes-Guridi,$^{1}$ J. Pachos,$^{
2}$ S. Bose,$^{3}$ V. Vedral,$^{^{1}}$ and S. Choi$^{4}$}
\address{$^{1}$Optics Section, The Blackett
Laboratory,
Imperial College, London SW7 2BZ, United Kingdom \\
$^{2}$Max-Planck-Institut f\"ur Quantenoptik, D-85748 Garching,
Germany \\
$^{3}$Center for Quantum Computation, Clarendon Laboratory,
    University of Oxford,
    Parks Road,
    Oxford OX1 3PU, England \\
$^{4}$Department of Chemistry,
    University of Rochester, Box 270216
    Rochester, New York 14627-0216}

\maketitle
\begin{abstract}
We propose a possible scheme for generating spin-$J$ geometric
phases using a coupled two-mode Bose-Einstein condensate (BEC).
First we show how to observe the standard Berry phase using Raman
coupling between two hyperfine states of the BEC. We find that the
presence of intrinsic interatomic collisions creates degeneracy in
energy that allows implementation of the non-Abelian geometric
phases as well. The evolutions produced can be used to produce
interference between different atomic species with high numbers of
atoms or to fine control the difference in atoms between the two
species. Finally, we show that errors in the standard Berry phase
due to elastic collisions may be corrected by controlling
inelastic collisions between atoms.
\end{abstract}

\begin{multicols}{2}
\section{INTRODUCTION}
    Geometrical phases in quantum theory have
attracted considerable interest since Berry \cite{berry} showed
that the state of a system undergoing an adiabatic and cyclic
evolution acquires this purely geometric feature in addition to
the usual dynamical phase. The term "geometric" comes from the
fact that the phase factor acquired by the state depends only on
the path followed by the state but not on the rate at which it is
traversed. If the system is nondegenerate, the geometric phase is
simply a complex number called the Abelian phase, but in general
it is a unitary matrix inducing transitions between degenerate
states, which is called the non-Abelian phase or holonomy
\cite{hol}. Geometric phases have been proposed \cite{prop} and
tested \cite{tom} in a variety of settings. Generalizations of
Berry's phase analysis to nonadiabatic, noncyclic, and nonunitary
evolutions have also been achieved \cite{gen}. Sj\"oqvist {\it et.
al.} proposed an operationally well defined generalization of the
geometric phase for mixed states \cite{sjoq} and more recently, a
fully quantized version of the phase has been given that considers
vacuum induced effects \cite{vacuum}.
%%%%%% SC [Swapped the order of sentences below -- more logical]
Holonomies and Berry phases have relevance in implementing quantum
computation \cite{jiannis}, where a universal set of quantum gates
can be performed in a fault-tolerant way by a succession of
geometrical unitaries \cite{Ellinas}. Moreover, geometric
evolution of states may also have importance in manipulating
quantum systems such as Bose-Einstein condensates (BECs). For
example, a certain type of geometric phase have been used to
create vortices in BECs \cite{becberry}. In this paper, we propose
a method of testing both Abelian and non-Abelian geometric phases
for a general spin-$J$ system modeled by two coupled Bose-Einstein
condensates. This is interesting, because it allows us to test
geometric phases for spin values as mesoscopic as $J\sim 10^4$, a
feat not accomplished yet in any other system.

Trapped atomic BECs, first achieved some seven years ago,
%%% SC
provide us with the ability to make mesoscopic quantum objects
containing of the order 10$^{6}$ atoms in the same quantum state.
Although a mesoscopic system may, in principle, be constructed
using photons as well, a crucial distinction is that atoms may be
stored for longer times. A longer storage time (of the order of an
hour for magnetic traps \cite{libbrecht}) implies a longer
decoherence time scale of quantum states, thereby aiding the
implementation of adiabatic evolution (required for the Berry
phase). Geometrical phases are generated by adiabatically varying
the Hamiltonian of a system in a cyclic fashion. This can be done
in BECs since the states of ultracold atomic samples can be
manipulated by electromagnetic fields.
%%%%%% SC [Moved this sentence up from below -- more logical]

  The Hamiltonian describing a fixed number of atoms
in two different internal levels trapped by a magnetic potential
can be approximated, for systems composed of a few thousand atoms,
by a two-mode Hamiltonian \cite{Steel,cirac}. The Schwinger
oscillator model allows to conveniently express the two-mode
problem in terms of angular momentum operators. In this way the
Abelian and non-Abelian phases generated are associated with
spin-$J$ states, where the spin is related to the total number of
atoms in the condensate.  We use two-photon excitations, which
generate a coherent superposition in the two-mode BEC, to
demonstrate the Berry phase when the collisions between atoms are
neglected. We then extend this idea further to consider the
possibility of measuring holonomies including the nonlinear term
due to collisional interaction between particles for which the
Hamiltonian is degenerate. To produce Abelian and non-Abelian
transformations when the non-linear term in the Hamiltonian is
considered, in addition to the two-photon excitations, inelastic
collisions must be manipulated. Together the two-photon
excitations and the inelastic collisions correspond to a two-mode
displacement of the eigenstates of the system and this may have
potential applications in manipulating BECs. We shall show that,
if collisions between atoms are seen as errors in the Berry phase
generated by the linear Hamiltonian, these errors can be corrected
by introducing inelastic collisions.

The paper is organized as follows. In Sec. II we describe, using
angular momentum operators, the physical system we consider for
generating geometric phases: the two-mode BEC. Section III is
devoted to explain how Berry phases and holonomies arise,
presenting the main mathematical formalism used through the paper.
In section IV we discuss how to generate Berry phase in the BEC
described in section II
%SC changed I to II
 and we propose a scheme for detecting this phase. In
Sec. V we move to consider the holonomies related to the two-mode
BEC. Section VI contains final remarks concerning the generation
of the Berry phase when collisions between the atoms in the
condensates cannot be neglected, and finally the paper concludes
in Sec. VII.
%SC changed VI to VII

\section{THE TWO-MODE BOSE-EINSTEIN CONDENSATE}
    A physical realization of our system is two
condensates in different hyperfine
levels $|A\rangle$ and $|B\rangle$, such as those
already produced
%%%%%%%%% Altered by SC
in experiments by Myatt {\it et. al} \cite{JILA2} and Stenger {\it
et. al} \cite{MIT}. In the Rb system of Ref.\cite{JILA2}, an
external laser is applied to induce a Josephson-like coupling and
the detuning of the laser is adiabatically changed to produce
various transitions. Alternatively, for the Na spinor system of
Ref. \cite{MIT}, state-dependent magnetic field gradient may be
applied to induce Josephson tunneling.
%%SC slight rearrangement of sentences
The Hamiltonian for the system can be written under the two-mode
approximation, taking the annihilation operators to be $a$ and $b$
for the two distinct hyperfine states:
%%%%%%%%%%

\begin{eqnarray}
H & = & H_{a}+H_{b}+H_{int}+H_{las},\\
H_{a}& = &
\omega_{a}a^{\dagger}a+\frac{U_{a}}{2}a^{\dagger}
a^{\dagger}aa,\\
H_{b}& = &
\omega_{b}b^{\dagger}b+\frac{U_{b}}{2}b^{\dagger}b^{\dagger}bb,
\\H_{int} & = &
\frac{U_{ab}}{2}a^{\dagger}ab^{\dagger}b,
\\H_{las} & = & -\lambda(a^{\dagger}be^{-i\Delta
t}+b^{\dagger}ae^{i\Delta t}),
\end{eqnarray}
where $H_{a}$ and $H_{b}$ describe the two condensates undergoing
self-interactions and $H_{int}$ and $H_{las}$ describe the
condensates interacting with one another via collisional and
laser-induced interactions, respectively. $\Delta$ is the detuning
of the laser from the $|A\rangle \rightarrow |B\rangle$
transition. We note in particular that $H_{las}$ describes
Josephson-like coupling which interchanges internal atomic states
in a coherent manner.

As discussed above, the Hamiltonian can be written in a more
suitable way by
%SC
employing the Schwinger angular momentum [$SU(2)$] operators
defined as $J_{x}=\frac{1}{2}(a^{\dagger}b+ ab^{\dagger}),
J_{y}=(1/2i)(a^{\dagger}b-ab^{\dagger}), and
J_{z}=\frac{1}{2}(a^{\dagger}a-b^{\dagger}b)$. The Casimir
invariant $J^2=J_x^2 +J_y^2 +J_z^2$ has eigenvalues $j(j+1)$ and
$j$ represents the total number $2N=2(N_A+N_B)$ of the two
different species of atoms ($N$ could be $10^4$ for the type of
condensates we consider). In terms of Schwinger operators, the
Hamiltonian takes the simple form
\begin{equation} \label{eq:comham}
H=\alpha J_{z}+\beta
J_{z}^{2}+\gamma[\cos(\phi)J_{x}+\sin(\phi)J_{y}]
\end{equation}
where $\alpha = \omega_{a} - \omega_{b} + (2J-1)(U_{a} -
U_{b})/2$, $\beta = (U_{a} + U_{b} - U_{ab})/2$, and $\phi=\Delta
t$. The eigenvalues $m$ of the operator $J_{z}$ represents the
difference $2(N_A-N_B)$ in the number of atoms in different
hyperfine levels, while $J_{x}$ (and $J_{y}$) takes on the meaning
of the relative phase between the two species. It is noted here
that $s$-wave scattering lengths may be tuned using Feshbach
resonances by the application of an external magnetic
field\cite{inouye}. The factors $U_{a}$, $U_{b}$, $U_{ab}$, and
consequently $\alpha$ and $\beta$ are therefore adjustable
parameters giving us important additional degrees of freedom.
%SC
The above Hamiltonian [Eq.(\ref{eq:comham})] is sufficient to
generate the Abelian Berry phase if the collision term ($\beta
J_{z}^{2}$) is negligible. We shall describe later how terms such
as $J_x J_y$ and $J_x^2$ (required for the non-Abelian geometric
phase) can be added to the above Hamiltonian
 by inducing inelastic collisions.

\section{Geometric Evolutions}
 We will now describe briefly how geometric
phases arise. Holonomies are unitary transformations of geometric
origin generated by varying a set of external parameters
$\sigma=\{\sigma_\mu\}_{\mu=0}^{k}$ featuring in the Hamiltonian
of a system, in a cyclic and adiabatic way. Their dimensionality
$n$ equals the degree of degeneracy of the eigenspace. The Berry
phase is the special case when the eigenspace is nondegenerate,
and the unitary transformation is then one dimensional, i.e., a
complex number. Consider that the adiabatic variation of the
Hamiltonian is given by $H(\sigma)= {\cal U}(\sigma)H_{0} {\cal
U}^{\dagger}(\sigma)$ (where the parameters $\sigma$ vary on some
control manifold $\mathcal{M}$) with the ${\cal U}(\sigma)$
transformation  being such that the degeneracy structure of the
initial Hamiltonian $H_{0}$ is preserved. For this purpose, the
parameters must be varied slowly with respect to any time scale
associated with the system dynamics. After $\sigma$ completes a
loop $C$ in $\mathcal{M}$, an initially prepared state
$|\Psi_{in}\rangle$ is mapped to $|\Psi_{out}\rangle=e^{-i E T}
\Gamma_{A}(C)|\Psi_{in}\rangle$ where $T$ is the overall time of
the evolution, and $E$ is the energy of the degenerate space where
$|\Psi_{in}\rangle$ and $|\Psi_{out}\rangle$ belong.
$\Gamma_{A}(C)$ (termed the holonomy associated with non-Abelian
connection forms) appears due to the nontrivial topology structure
of the degenerate space and is given by
\begin{equation}
\Gamma_{A}(C)={\bf P}\exp{\int_{C}A} \,\, .
\label{geo}
\end{equation}
$\bf{P}$ is the path ordering symbol and the
Wilczek-Zee connection
$A$ \cite{hol} is defined as
\begin{equation}
A^{\mu}_{ij}:\equiv\langle
i|{\cal
U}^{\dagger}(\sigma)\frac{\partial}{\partial\sigma_{\mu}}{\cal
  U}(\sigma)|j\rangle
  \label{pot}
\end{equation}
for $i,j=1,...,n$ parametrizing states belonging to the same
degenerate eigenspace of $H_{0}$.

An alternative interpretation of the holonomy, which is needed
here to calculate explicitly the integral in Eq. (\ref{geo}), is
that
 the holonomy can be seen as the
exponential of the flux of the field strength
$F_{\sigma
  \tau}(\sigma,\tau)=-\p_\sigma A_\tau +\p_\tau
A_\sigma+[A_\sigma , A_\tau]$, through a surface parametrized by
$\sigma$ and $\tau$. If we consider the surface to be a
rectangular loop $C$ in the plane $(\sigma, \tau)$ with ordered
sides $1$, $2$, $3$ and $4$ , and represent the path order
exponential integrals of the connection for each side by $W_i$ for
$i=1,...,4$, for $T^{-1}(\sigma,\tau)=W_4 W_3$ we can state the
non-Abelian Stokes theorem \cite{Karp} as \bq && \Gamma_A(C)= W_4
W_3 W_2 W_1= {\bf P}_\tau e^{\int_{\Sigma(C)} T^{-1}F_{\sigma
    \tau}T d\sigma d\tau} \,\, ,
\label{stokes} \eq where ${\bf P}_\tau$ is the path ordering
symbol with respect to the $\tau$ variable only, unlike the usual
path ordering symbol {\bf P}, which is with respect to both
variables $\sigma$ and $\tau$. The evaluation of the holonomies
here is performed by the application of (\ref{stokes}).

\section{SPIN J BERRY PHASE}
\begin{figure}
\begin{center}
\leavevmode \epsfxsize=8cm \epsfbox{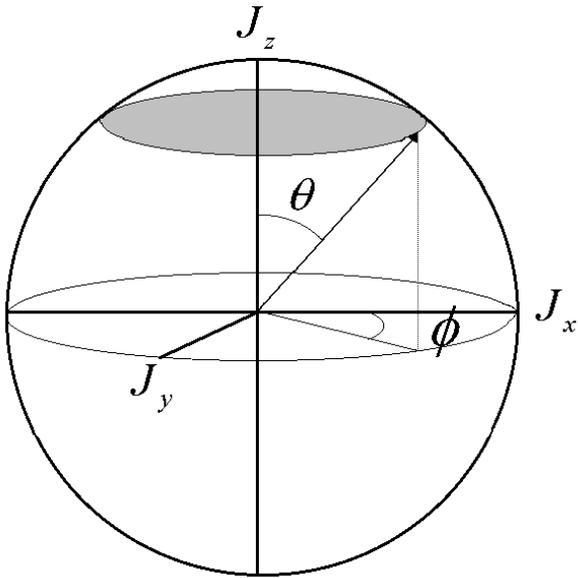} \caption{The spin-J
Berry phase is obtained by rotating the eigenvectors of $H_{0}$
first through an angle $\theta$ in the $xz$ plane and then through
an angle $\phi$ in the $xy$ plane. The phase is proportional to
the area traversed by the vector.} \label{bbec}
\end{center}
\end{figure}

To generate a Berry phase in the BEC, let us first ignore the
terms in the Hamiltonian due to collisions.
%%%%%%%%%%% Altered by SC
A BEC can be described by such a Hamiltonian when one may assume
that the condensate is dilute enough that the collisional
interactions between the atoms (both inter- and intraspecies)
become insignificant compared to the coupling rate generated by
the strong external field. In principle, one may also consider the
case where through the Feshbach resonance $U_{a} + U_{b} = U_{ab}$
($\beta = 0$), or one can assume $U_{ab}= U_{a} = U_{b}$, and tune
the scattering length of either one of the species so that either
$U_{a}$ or $U_{b}$ is reduced down to zero energy\cite{verhaar}.
Experimentally, one needs to ensure that losses due to three-body
recombinations are minimized.
%%%%%%%%%%%%
In this case the nonlinear term in the Hamiltonian vanishes so
that
\begin{equation}
H=\alpha J_{z}+\gamma[\cos(\phi)J_{x}+\sin(\phi)J_{y}].
\end{equation}
This Hamiltonian corresponds to the motion of a spin $J$ particle
in a magnetic field with amplitude $B=\sqrt{\alpha^2+\gamma^2}$,
whose direction is slowly varying as $\phi$ changes. $\phi=\Delta
t$ can be made to vary arbitrarily slowly (so that the adiabatic
approximation holds true) by choosing arbitrarily small detuning
$\Delta$. For a fixed $\gamma$, varying $\phi$ over a complete
loop produces a nontrivial Berry phase. We now proceed to
calculate this. The Hamiltonian can be written as
\begin{equation}
H(\phi,\theta)={\cal U}(\phi,\theta) H_{0} {\cal
U}^{\dagger}(\phi,\theta)
\end{equation}
where $H_{0}=\alpha_0 J_{z}$ and ${\cal
U}(\phi,\theta)=\exp(-i\phi J_{z})\exp(-i\theta J_{y})$ with
$\alpha=\alpha_0\cos{\theta}$ and $\gamma=\alpha_0\sin{\theta}$.
This unitary transformation on $H_0$ corresponds to a two-mode
displacement with amplitude $\theta/2$ and phase $\phi$. The
eigenvectors of $H_{0}$ are first rotated through an angle
$\theta$ on the $xz$ plane and then rotated through an angle
$\phi$ in the $xy$ plane (as shown in Fig.\ref{bbec}) to obtain
the eigenvectors of $H$. The eigenstates of $H_{0}$ are the usual
angular momentum eigenstates with $J_{z}|j, m\rangle=m|j,
m\rangle$. This implies that the eigenstates of our original
Hamiltonian $H$ are $|\psi\rangle={\cal U}(\phi,\theta)|j,
m\rangle$ and we can then calculate the Berry phase using
expressions (\ref{geo}) and (\ref{pot}) for $i=j=m$. The
connection components generated by the transformation ${\cal
U}(\phi,\theta)$ for the eigenstates of $H_{0}$ are given by
\begin{eqnarray}
A_{\phi}&=&\langle
j,m|U^{\dagger}(\phi,\theta)\frac{\partial}{\partial \phi
}U(\phi,\theta)|j,m\rangle\nonumber\\&=&-i\langle
j,m|(\sin(\theta) J_{x} +\cos(\theta)J_{z})|j,m\rangle
\nonumber\\&=&-im\cos{\theta}
\end{eqnarray}
and
\begin{eqnarray}
A_{\theta}&=&\langle
j,m|U^{\dagger}(\phi,\theta)\frac{\partial}{\partial \theta
}U(\phi,\theta)|j,m\rangle\nonumber\\&=&-i\langle j,m|
J_{y}|j,m\rangle=0.\end{eqnarray}

The field strength is $F_{\phi,\theta}=im\sin\theta$ and we find
that, if we vary $\theta$ from zero to a fixed value and $\phi$
from $0$ to $2\pi$, the state $|j,m\rangle$ acquires a phase
\begin{equation}
\label{bp}
\phi^{m}_{Berry}=i\int_{0}^{2\pi}\int_{0}^{\theta}F_{\phi,\theta^{\prime}}d\theta^{\prime}
d\phi=-2\pi m(1-\cos{\theta}) \,\, ,
\end{equation}
which is $m$ times the solid angle subtended by the circuit at the
origin in parameter space. During the evolution, the state also
acquires a dynamical phase $\gamma^{m}_{dyn}=\int_{0}^{T}\langle
j,m|H_{0}|j,m\rangle=m \alpha_0 T$. This phase can be eliminated
by choosing an adequate evolution time $T$ for which the dynamical
phase is a multiple of $2\pi$. Note that the Berry phase does not
depend on $T$ but only on the geometry of the loop $C$.

%%%%%%%%%%%%%%%%%%%%%%%%%%%%%%%%%%%%%%%%%%

Now we shall discuss a scheme for detecting the Berry phase. We
first prepare the system in the state $|j,j\rangle$, which, when
represented in terms of the population of the two modes, is simply
$|N\ran\otimes|0\rangle$.  We then switch on laser fields with
detuning $\Delta=0$, and vary $\alpha$ and $\gamma$ slowly in such
a way that $|j,j\rangle$ (following the Hamiltonian adiabatically)
evolves to ${\cal U}(0,\pi/2)|j, j\rangle$. Now we implement an
adiabatic loop of the Hamiltonian in the parameter space given by
the transformation ${\cal U}^{\dagger}(0,\theta){\cal
U}(2\pi,0){\cal U}(0,\theta)$, with $\cos{\theta}=1/2$. The middle
transformation  ${\cal U}(2\pi,0)$ is obtained by switching on the
detuning $\Delta$ and letting the Raman transition on for a time
such that $t=\frac{2\pi}{\Delta}$). If we choose the time of the
loop of the Hamiltonian such that the dynamical phase is
eliminated (i.e., a multiple of $2\pi$) for all states, then the
evolution of the state during this loop will be purely due to
geometric phases. Next, the transformation ${\cal
U}^{\dagger}(0,\pi/2)$ is applied to the state. For our choice of
$\theta$, the Berry phase will be such that the final state after
all transformations will be orthogonal to $|j,j\rangle$ . The
presence of a Berry phase can now be verified by measuring the
population of the second mode, which will now always be nonzero.
This method is a generalization of the usual
Hadamard-Berry-Hadamard method used for the detection of the
spin-$1/2$ Berry phase.
%%%%%%%%%%%%%%%%%%%%%%%%%%%%%%%%%%%%%

\section{HOLONOMIC EVOLUTION IN THE TWO-MODE BEC}
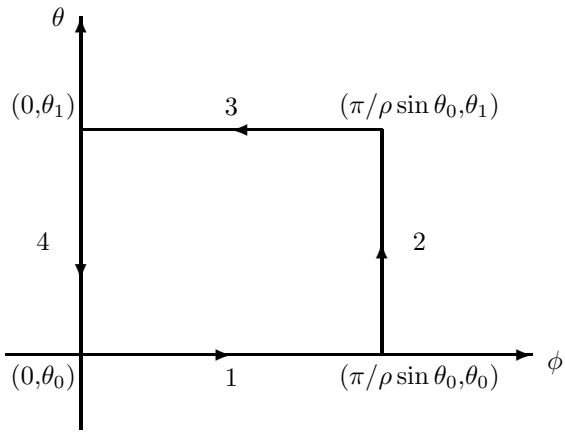
\begin{figure}
\setlength{\unitlength}{1cm}
\begin{picture}(6,6)\thicklines
%\put(1,1){\framebox(4,3)[4,4]{}}
\put(0,1){\vector(1,0){3}} \put(3,4){\line(-1,0){2}}
\put(5,4){\line(0,-1){2}} \put(0,1){\vector(1,0){7}}
\put(1,0){\vector(0,1){5.5}} \put(1,4){\vector(0,-1){2}}
\put(5,4){\vector(-1,0){2}} \put(5,1){\vector(0,1){1.5}}
\put(0,0.2){\makebox(1,1){(0,$\theta_{0}$)}}
\put(0,3.8){\makebox(1,1){(0,$\theta_{1}$)}}
\put(0.2,5){\makebox(1,1){$\theta$}} \put(0,2){\makebox(1,1){4}}
\put(5,3.8){\makebox(1,1){($\pi/\rho\sin\theta_{0}$,$\theta_{1}$)}}
\put(5,0.2){\makebox(1,1){($\pi/\rho\sin\theta_{0}$,$\theta_{0}$)}}
\put(6.8,0.4){\makebox(1,1){$\phi$}} \put(5,2){\makebox(1,1){2}}
\put(2.5,0.2){\makebox(1,1){1}} \put(2.5,3.8){\makebox(1,1){3}}
\end{picture}
\caption{The loop C for the non-Abelian Stokes theorem.}
\label{fig:path}
\end{figure}

A geometric evolution in the condensate becomes a more complicated
transformation when collisions are considered. The nonlinear term
in the Hamiltonian allows for the possibility of degeneracy and,
by slowly varying the Hamiltonian through the parameters, we can
then generate holonomies. For performing a holonomic evolution the
state of the system must be confined to a degenerate subspace at
all times. A two-dimensional
 degenerate subspace can be created by making two
of the eigenvalues of the Hamiltonian $H_0=\alpha_0 J_{z}+\beta_0
J_{z}^{2}$ equal. Choosing $\alpha_0 / \beta_0= -(2m+1)$, the
states $|j,m\rangle$ and $|j,m+1\rangle$ have the same energy. We
shall transform the nonlinear Hamiltonian with the same unitary
transformation that we used for the linear one, ${\cal
U}(\phi,\theta)=\exp(-i\phi J_{z})\exp(-i\theta J_y)$. This will
result in a transformed Hamiltonian that includes inelastic
collisions, but the states are transformed in the same manner as
in the linear case. In particular, we obtain $ U(\theta) H_0
U^\dagger(\theta)= \alpha_0 [\sin(\theta)J_x+\cos(\theta)J_z)]
+\beta_0 [\sin ^2(\theta)J_x ^2 +\cos ^2(\theta)J_z ^2
+\sin(\theta)\cos(\theta)(2 J_x J_z +1)]$. The term $J_x J_z$
includes inelastic collisions of the form $a^\dagger b a^\dagger
a$, while $J_x^2$ includes terms of the form $\left. a^\dagger
\right.^2 b^2$.
%%%%%%%%%%%% Added by SC
Such inelastic collisional terms describe processes in which atoms
exchange the hyperfine states during a
 collision. We do not have to introduce such terms
artificially in our system as such processes are normally present
and only deliberately suppressed when one tries to produce a BEC.
The reason for this is that the atoms are generally confined in a
magnetic trap and such spin flips result in the loss of the atoms.
However, by using an optical confinement of the BEC\cite{OptTrap}
such problems are removed, as the optical dipole force is not
selective about the hyperfine states of the atoms. The measures to
suppress inelastic collision can be removed with an optical trap
and, indeed, one may even enhance these processes by inducing
suitable Zeeman shifts. This is possible provided the total
angular momentum and energy are conserved on collision. By freeing
the channel through which excess angular momentum is translated
into an overall relative rotational motion of the colliding atoms,
the inelastic collisional processes can be enhanced.
%%%%%%%%%%%%
By taking $\theta$ small it is possible to meet the
experimental
values for the production ratios of those terms in a
two-mode
BEC. The connection components generated by the
transformation
${\cal U}(\phi,\theta)$ and for the above degenerate
states are
given by
\[ \begin{array}{cc}
A_\phi=
 i \left[  \begin{array}{ccc}   -m\cos{\theta} & {\rho
\over 2}\sin{\theta} \\
                                {\rho \over
2}\sin{\theta} & -(m+1)\cos{\theta} \\
\end{array} \right] ,
\,\, A_\theta= {\rho \over 2}
 \left[  \begin{array}{ccc}     0 & 1 \\
                               -1 & 0 \\
\end{array} \right] ,
\end{array}\]
where $\rho=\sqrt{(j-m)(j+m+1)}$. We now consider the case $m
\approx 0$ which corresponds to almost equal numbers of particles
in both condensates. In this case $\rho \approx \sqrt{j(j+1)}$.
For a large number of atoms $\rho\gg 1$, we can neglect terms that
are small compared to $\rho$ in the following analysis. As the
connection components $A_\phi$ and $A_\theta$ do not commute with
each other we have to employ the non-Abelian Stokes theorem to
evaluate the holonomy $\Gamma_A(C)$. Indeed, following the
procedure presented in \cite{Karp,Pachos}, for a rectangular loop
$C$ with vertex coordinates $\{(0,\theta_0),(\phi=\pi/(\rho \sin
\theta_0),\theta_0),(\phi=\pi/(\rho \sin
\theta_0),\theta_1),(0,\theta_1)\}$ (shown in Fig.\ref{fig:path})
we obtain the holonomy \bq && \Gamma_A (C) = \no \no && \exp
\left\{ - i(\cos{\theta_{1}}-\cos{\theta_{0}}) \left[(m+{1 \over
2}){\bf 1}-{\rho \over \sin{\theta_{0}} }\,\hat \sigma_2\right]
\right\} \nonumber \eq where $\hat \sigma_2$ is the Pauli matrix.
To obtain the above result we have used the approximation that for
$\theta_0$ large compared to the variation $\theta_1-\theta_0$,
the function $\sin(\theta)$ is almost constant compared to $\sin
(\rho \, \theta)$. As an application for $\theta_1-\theta_0=
\pi/(2 \rho)$ we can obtain a change of state from $|j,m\rangle$
to $|j,m+1\rangle$ (transfer of two atoms from one mode to the
other). In addition, starting with the atom number state
$|N_A\ran\otimes|N_B\ran$ and for $\theta_1-\theta_0= \pi/(4
\rho)$, one can produce up to an overall phase the state
$(|N_A\ran \otimes
|N_B\ran-|N_A+2\ran\otimes|N_B-2\ran)/\sqrt{2}$. This interference
procedure can be used, e.g., in high-precision measurements for
the construction of quantum gyroscopes \cite{Delgado}.

\section{FINAL REMARKS}
The standard Berry phase can also be generated when collisions are
included. By relaxing the degeneracy condition $\alpha_0 /\beta_0
=-(2 m+1)$ and performing the same transformation ${\cal
U}(\phi,\theta)$ to the Hamiltonian the same Berry phase as in Eq.
(\ref{bp}) is generated even in the presence of collisions. This
is, however, true only when both elastic and inelastic collisions
are considered. One can think of this in the following way:
Considering elastic collisions to be errors, in order to generate
the same Berry phase as in the collision-free Hamiltonian, there
must be a correction achieved by including inelastic terms.

Finally, we shall give an analytic expression for the unitary
transformation that produces from the nonlinear Hamiltonian $H_0$
the term $\gamma J_x$ for small $\gamma$, but arbitrary $\alpha_0$
and $\beta_0$. The only condition needed in order for the
perturbation expansion to be valid is that the Hamiltonian should
not be degenerate, i.e., $\alpha_0/\beta_0$ should not be an
integer. Indeed, with easy algebraic steps we have
\begin{equation}
H \approx U(\phi)U(\gamma) H_0 U^\dagger(\gamma)U^\dagger(\phi) \,\, ,
\end{equation}
where the $\phi$ dependence is exact, while the $\gamma$
dependence is valid for weak Josephson-like coupling, $\gamma\ll
1$. The unitaries are given by $U(\phi)=\exp(i\phi J_z)$ and
$U(\gamma)=\exp(i\gamma G)$ with
\begin{eqnarray}
G &&:= \sum _{k+l=0}^\infty a_{kl} \left.J_z \right.^k
J_y \left.
J_z\right.^l\\ \nonumber a_{kl}&=& (-1)^{k+l+1}
{(k+l)! \over k!
\, l!} { \beta_0^{k+l} \over
  \alpha_0^{k+l+1}} \,\, .
\label{ttt}
\end{eqnarray}
$U(\phi)$ and $U(\gamma)$ rotate the basis of eigenvectors of
$H_0$. As $G$ is evaluated only to first order in $\gamma$ it is
impossible to evaluate from it a nonvanishing Berry phase, as its
calculation involves two exterior derivatives of the
transformation $U(\gamma)$.

\section{CONCLUSIONS}
 We have presented a procedure for evolving
the state of a two-mode BEC in a geometrical fashion. The BEC
setup described here due to its two mode bosonic nature resembles
other bosonic models studied in connection with Berry phases and
holonomies \cite{Rabei}. Here we have focused on the spin-$J$
description of the condensate that is apparent during the
measuring procedure of the Berry phase. In particular, in the
limit where collisions can be neglected the state of the
condensate acquires a Berry phase by varying the displacement
parameter in a cyclic and adiabatic way. This resembles a spin-$J$
particle, where $J$ is mesoscopic. Berry's phase is manifested by
varying the direction of the magnetic field. Considering the
degeneracy introduced by the collisions between atoms, a holonomic
evolution, generated by slowly changing the parameters, allows for
controlled transfer of population between modes. In addition to
allowing for tests of Abelian and non-Abelian geometrical phases
for mesoscopic $J$, this might also be useful as a procedure for
manipulating the condensate.
\begin{acknowledgments}
We thank Jesus Rogel-Salazar and Janne Ruostekoski for valuable
discussions. This research has been partly supported by the
European Union (contract IST-1999-11053), EPSRC and
Hewlett-Packard. I.F.-G. would like to acknowledge Consejo
Nacional de Ciencia y Tecnologia (Mexico) Grant no. 115569/135963
for financial support.
\end{acknowledgments}

\end{multicols}{2}

\end{document}